\documentclass[aps,prl,reprint,superscriptaddress,nofootinbib,floatfix]{revtex4-2}
\usepackage{amsmath,amssymb,bm,graphicx,mathtools,hyperref,xcolor}
\hypersetup{colorlinks=true,linkcolor=blue,citecolor=blue,urlcolor=blue}
\newcommand{\Order}{\mathcal{O}}
\newcommand{\ee}{\mathrm{e}}
\newcommand{\theorembox}[1]{%
  \begin{center}%
  \fcolorbox{black!35}{black!8}{\parbox{0.96\columnwidth}{\small #1}}%
  \end{center}}
\usepackage{url}
\usepackage{fontawesome5}
\usepackage{scalerel}
\usepackage{tikz}
\usetikzlibrary{svg.path}
\definecolor{orcidlogocol}{HTML}{A6CE39}
\tikzset{
  orcidlogo/.pic={
    \fill[orcidlogocol] svg{M256,128c0,70.7-57.3,128-128,128C57.3,256,0,198.7,0,128C0,57.3,57.3,0,128,0C198.7,0,256,57.3,256,128z};
    \fill[white] svg{M86.3,186.2H70.9V79.1h15.4v48.4V186.2z}
                 svg{M108.9,79.1h41.6c39.6,0,57,28.3,57,53.6c0,27.5-21.5,53.6-56.8,53.6h-41.8V79.1z M124.3,172.4h24.5c34.9,0,42.9-26.5,42.9-39.7c0-21.5-13.7-39.7-43.7-39.7h-23.7V172.4z}
                 svg{M88.7,56.8c0,5.5-4.5,10.1-10.1,10.1c-5.6,0-10.1-4.6-10.1-10.1c0-5.6,4.5-10.1,10.1-10.1C84.2,46.7,88.7,51.3,88.7,56.8z};
  }
}
\newcommand\orcidicon[1]{\href{https://orcid.org/#1}{\mbox{\scalerel*{
\begin{tikzpicture}[yscale=-1,transform shape]
\pic{orcidlogo};
\end{tikzpicture}
}{|}}}}

\definecolor{mycolor}{RGB}{0,0,204}
\definecolor{pink}{RGB}{255,0,127}

\usepackage{tabularx}

\newcommand{\SM}{\textsuperscript{SM}}

\begin{document}

\title{Controlled Penumbral Inflation from Monodromic Valleys}

\author{Pirzada \orcidicon{0009-0002-2274-9218}}
\email{pirzada@itp.ac.cn}
\affiliation{CAS Key Laboratory of Theoretical Physics, Institute of Theoretical Physics, Chinese Academy of Sciences, Beijing 100190, China}
\affiliation{School of Physical Sciences, University of Chinese Academy of Sciences, No. 19A Yuquan Road, Beijing 100049, China}
\author{Tianjun Li}
\email{tli@itp.ac.cn}
\affiliation{School of Physics, Henan Normal University, Xinxiang 453007, P. R. China}

\begin{abstract}
Realizing controlled, single-clock inflation in string theory is fundamentally obstructed by the backreaction of heavy moduli. We show that in the \emph{penumbra}---the near-boundary regime of complex-structure moduli space where asymptotic symmetries are partially broken---this obstruction can be exactly quantified. We derive a covariant control theorem demonstrating that local branch data dictate whether a monodromic valley supports a controlled inflationary plateau, thereby isolating the first controlled penumbral inflationary window. The result turns the penumbra from a geometric regime into a dynamical filter. In the axion--saxion effective theory, a branch-displacing odd term generates a plateau when $\Delta\equiv p+2\nu-q>0$, while covariant single-clock control further requires $p<2$, or $p=2$ with $12A_pm^2/(dV_0)\gg1$ over the observational window. This splits penumbral valleys into no plateau, uncontrolled plateau, and controlled plateau before global completion is attempted. We identify a minimal analytic family with a closed-form valley and an invariant attractor equation for the full two-field dynamics, providing the first exactly solvable penumbral realization that remains predictive under the next penumbral order. The controlled corridor targets $r\sim10^{-3}$ with the correlated running $\alpha_s\simeq-r/2$ for the $d=q=1$ benchmark, providing a falsifiable target for LiteBIRD/CMB-S4.
\end{abstract}

\maketitle

{\textbf{Introduction.}--}Axion monodromy remains one of the clearest routes to parametrically long inflationary trajectories in string theory~\cite{Silverstein:2008sg,McAllister:2008hb,Kaloper:2008fb}. The decisive obstruction is no longer field range alone. A long valley may still fail because heavy-sector backreaction, canonical distance, or loss of adiabaticity destroys the single-clock description~\cite{Flauger:2009ab,Dong:2010in}. F-term constructions, hierarchy analyses, universal axion backreaction, four-form descriptions, and large-distance EFT studies established how monodromy reshapes potentials and exposes UV sensitivity~\cite{Marchesano:2014mla,Blumenhagen:2014nba,Hebecker:2014eua,Hebecker:2017lxm,Baume:2016psm,Valenzuela:2016yny,Landete:2018kft,Grimm:2020cda,Grimm:2018ohb,Lanza:2021qsu,Pajer:2024uvs}. Hyperbolic attractors explain why logarithmic canonical variables generate plateau-like observables~\cite{Kallosh:2013hoa,Carrasco:2015rva,Galante:2014ifa,Roest:2015qya,Kallosh:2015zsa}. 
The decisive need for top-down work is a discriminator: which long monodromic valleys are controlled inflationary targets near, though not at, an infinite-distance boundary?

\begin{figure}[ht!]
  \includegraphics[width=0.98\columnwidth]{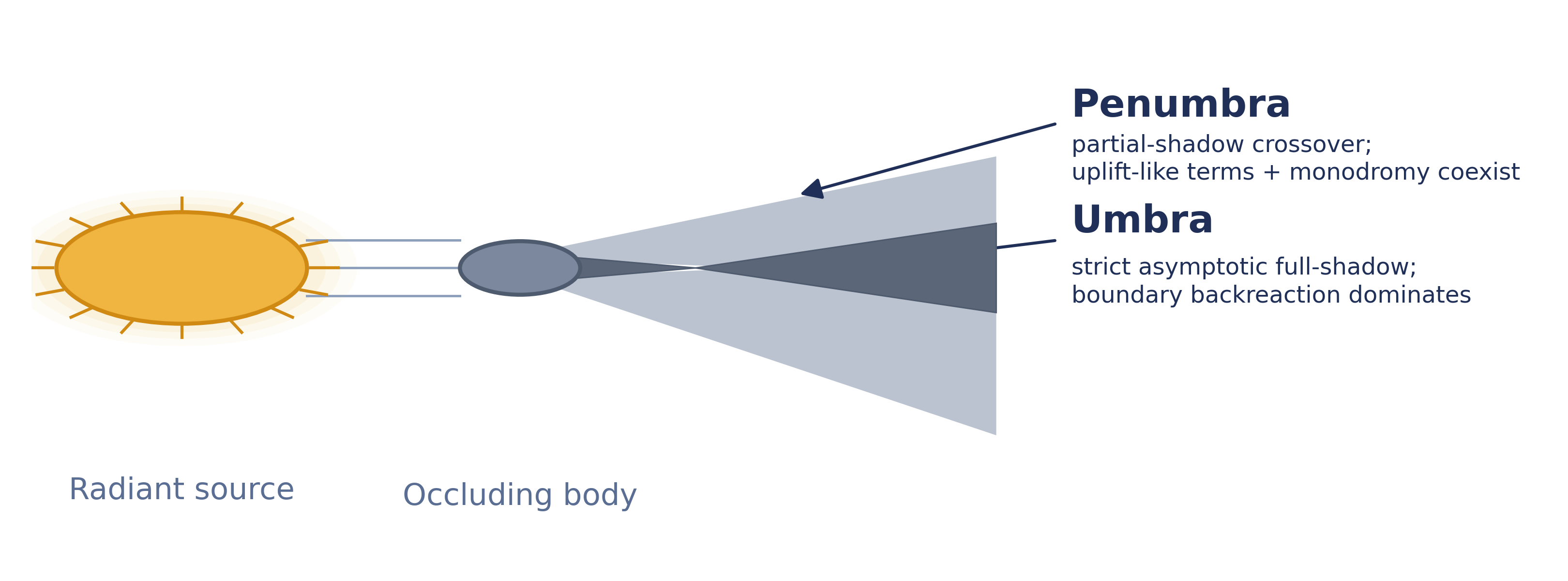}
  \caption{\textbf{Umbra and penumbra terminology.} The schematic fixes the language used throughout: the umbra is the strict asymptotic full-shadow regime, while the penumbra is the near-boundary partial-shadow crossover where uplift-like terms and polynomial monodromy still coexist.}
  \label{fig:umbra_penumbra}
\end{figure}

Recent work isolated the \emph{penumbra} of complex-structure moduli space as the crossover region where uplift-like positive terms coexist with surviving polynomial monodromic corrections and long axion valleys~\cite{Lanza:2024jxo}. Throughout this Letter, \emph{umbra} denotes the strict asymptotic full-shadow regime near an infinite-distance boundary, whereas \emph{penumbra} denotes the near-boundary partial-shadow regime where those terms still coexist; Fig.~\ref{fig:umbra_penumbra} fixes the terminology. The result of Refs.~\cite{Lanza:2024jxo} was not a controlled inflationary solution: their valleys exhibit monodromic flattening, but the slow-roll and single-clock obstructions remain. The decisive advance is therefore not another long valley, but a discriminator deciding when a penumbral branch is upgraded to controlled inflation. We identify that discriminator: a monodromy-preserving odd branch term displaces the heavy minimum, rotates the light trajectory saxionically, and simultaneously determines whether the orthogonal mode remains adiabatically heavy. Thus the central question is not whether a long penumbral valley exists, but whether it survives the local control test before the harder step of global embedding is attempted.

Here we derive that criterion. Near moduli-space boundaries, curved-field EFT is rich enough to reveal flattening, while global viability still requires full compactification data~\cite{Baume:2016psm,Valenzuela:2016yny,Hebecker:2017lxm,Grimm:2018ohb,Lanza:2021qsu,Grimm:2020cda,Pajer:2024uvs,Grimm:2021tame,Douglas:2024tameness}.Global completion of penumbral vacua is computationally expensive. We show that branch data alone provide a rigorous filter: a valley must pass the local control theorem before it merits the enormous effort of global embedding. Theorem violation is a lethal obstruction; theorem satisfaction is a necessary (and locally sufficient) condition for controlled inflation.  If branch data already decide which valleys deserve the harder step of global completion, top-down model building becomes a filtered search problem.  We obtain a covariant control theorem, exhibit a flux-compatible Hodge block with integer branch data, construct an analytic family whose attractor structure is visible directly in the full two-field equations, and show that the observable corridor survives the complete next penumbral order. The result is a decision principle for top-down multifield inflation: existence, control, and predictive stability are tested before global completion is attempted. Except where global completion is explicitly discussed, the analysis is within this one-modulus branch EFT.

{\textbf{Local EFT and Universality Class.}---} Write the complex modulus as $z=a+is$, where $a$ is the periodic axion and $s>0$ its saxionic partner. The one-modulus penumbral sector is described by the hyperbolic metric
\begin{equation}
\frac{{\cal L}_{\rm kin}}{\sqrt{-g}}=\frac{d}{4s^2}\big[(\partial a)^2+(\partial s)^2\big],
\qquad X\equiv e-ma,
\label{eq:metric}
\end{equation}
where $d>0$ is the boundary metric coefficient, $m$ the monodromy charge, and $e$ a constant branch-location parameter. The coordinate $X$ is the invariant heavy branch variable: motion in $a$ shifts the branch, while $s$ controls how strongly that branch sector feeds back into the low-energy potential.

Through quadratic order in $X$, the potential is written as
\begin{equation}
V=V_0-c_qV_0s^{-q}+A_ps^{-p}X^2+B_{p+\nu}s^{-(p+\nu)}X+\cdots.
\label{eq:Vgen}
\end{equation}
The four leading structures respectively control the uplift, saxionic falloff, heavy restoring force, and branch displacement. $V_0>0$ is the uplift-like positive piece, $c_qV_0s^{-q}$ sets the leading saxionic falloff of the plateau, $A_ps^{-p}X^2$ is the quadratic restoring operator in the heavy direction, and the odd term $B_{p+\nu}s^{-(p+\nu)}X$ displaces the heavy minimum away from $X=0$. The integers $q$, $p$, and $\nu$ control, respectively, the plateau exponent, the asymptotic softness of heavy stabilization, and the decay rate of the branch displacement. Without this odd displacement, the valley remains effectively axionic and the late-slope obstruction persists. The odd term is the geometric trigger: it shifts the heavy minimum, rotates the light trajectory into the saxionic direction, and converts monodromic backreaction into an exponentially flat plateau. Eq.~\eqref{eq:Vgen} is the branch expansion of a Type-IIB complex-structure flux potential near a one-parameter Calabi--Yau boundary: the flux potential is bilinear in monodromy-invariant axion polynomials with coefficients scaling as inverse powers of $s$, so a chosen branch contains an $X$-independent sector together with linear and quadratic heavy terms unless symmetry or flux tuning removes them~\cite{Grimm:2020cda,Grimm:2021tame,Grimm:2022tameness,Lanza:2024ml}\SM. Unless stated otherwise, all statements below refer to this one-modulus branch class, with $V_0>0$, $A_p>0$, $q\ge1$, $\nu\ge1$, positive kinetic metric, no competing correction at order $s^{-q}$, and only perturbative mixing with already-stabilized additional moduli. Strong mixing with the axio-dilaton or K\"ahler sector lies outside the one-modulus universality class and must be tested in the global embedding.

Minimizing with respect to $X$ gives
\begin{equation}
\begin{aligned}
X_v(s)&=-\frac{B_{p+\nu}}{2A_p}s^{-\nu}+\Order(s^{-\nu-1}),\\
\varphi&=\sqrt{\frac d2}\log s+\Order(s^{-2\nu-2}),
\end{aligned}
\label{eq:valley}
\end{equation}
so the valley is asymptotically tangent to the saxion rather than to the axion. The branch displacement therefore rotates the light direction. Substituting the valley back into Eq.~\eqref{eq:Vgen} yields
\begin{equation}
U(\varphi)=V_0-c_qV_0\ee^{-q\sqrt{2/d}\,\varphi}
-\frac{B_{p+\nu}^2}{4A_p}\ee^{-(p+2\nu)\sqrt{2/d}\,\varphi}+\cdots.
\label{eq:Uphi}
\end{equation}
Inverse powers of $s$ become exponentials because the canonically normalized light field is logarithmic. The plateau is the backreacted image of the monodromic valley itself. The saxionic branch converts the late-slope obstruction into a parametrically small plateau slope: earlier effectively axionic valleys can flatten while retaining
\begin{equation}
\gamma_{\rm late}\equiv \frac{|dV_{\rm val}/d\phi_{\rm val}|}{V_{\rm val}}=\Order(1),
\end{equation}
whereas the saxionic branch gives
\begin{equation}
\gamma_{\rm saxionic}\equiv\frac{|U_{,\varphi}|}{U}\simeq \frac{\sqrt{d/2}}{q\,N_*}
\simeq1.3\times10^{-2}
\end{equation}
for $d=q=1$ and $N_*=55$.

The leading $q$-universal plateau is obtained whenever
\begin{equation}
\Delta\equiv p+2\nu-q>0,
\label{eq:Delta}
\end{equation}
in which case, with $N_*$ the number of e-folds before the end of inflation,
\begin{equation}
 n_s=1-\frac{2}{N_*}+\Order(N_*^{-2}),
 \qquad
 r=\frac{4d}{q^2N_*^2}+\Order(N_*^{-3}).
\label{eq:nsr}
\end{equation}
Eq.~\eqref{eq:nsr} defines the $q$-universal penumbral class. The same data that generate the plateau also test whether the orthogonal sector remains heavy enough for the reduced one-field description\SM.
\begin{figure*}[ht]
  \includegraphics[width=0.40\textwidth]{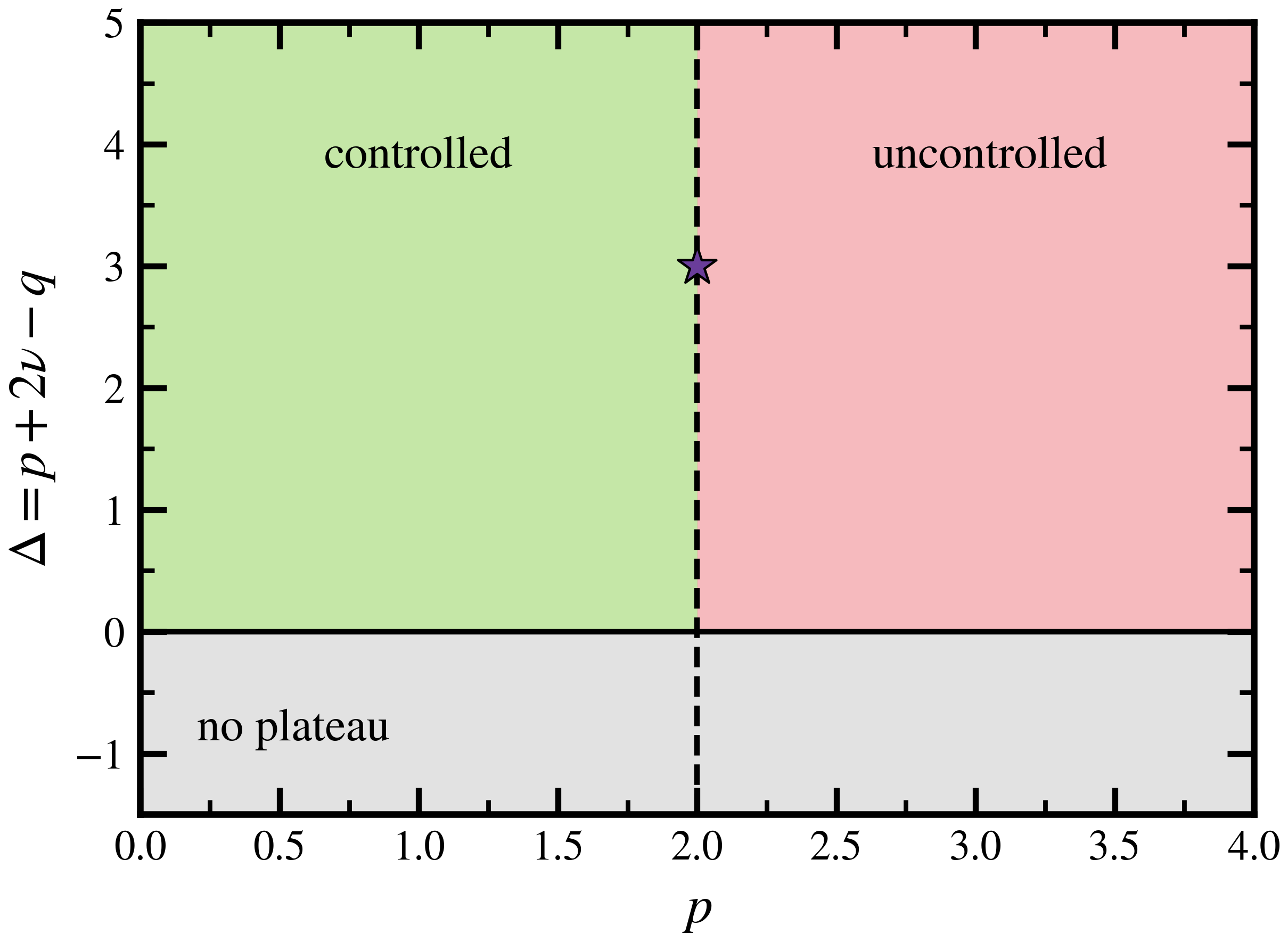}
  \includegraphics[width=0.40\textwidth]{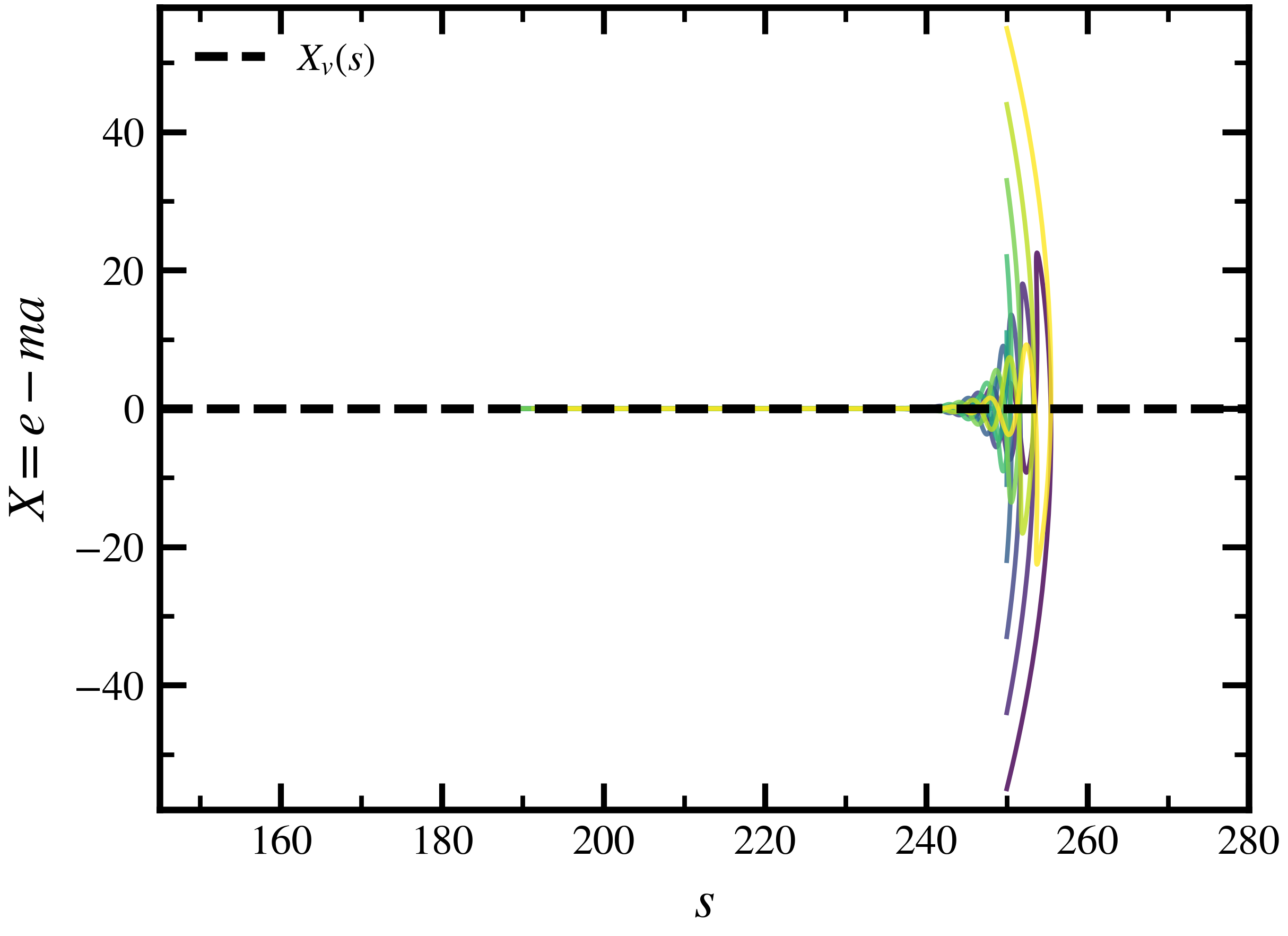}
  \caption{\textbf{Theorem and covariant control.} (a) Local theorem map in the $(p,\Delta)$ plane. The benchmark $(p,\Delta)=(2,3)$ sits on the asymptotic control boundary but inside the controlled region. (b)Representative two-field trajectories in $X=e-ma$ with the analytic valley. The restoring force acts on $X-X_v(s)$, driving the system to the valley before the cosmic-microwave-background (CMB) window ends.}
  \label{fig:control}
\end{figure*}

{\textbf{Control Theorem.}---} Adiabatic single-clock control is imposed covariantly: the orthogonal mode stays heavy, the turn rate stays small, and the reduced light trajectory captures the inflationary dynamics throughout the observational window. In curved field space the relevant quantity is the entropy mass~\cite{Gordon:2000hv,Senatore:2010wk,Achucarro:2012yr,Achucarro:2012sm,RenauxPetel:2015mga}, the basis-independent mass of fluctuations orthogonal to the trajectory,
\begin{equation}
m_s^2\equiv m_\perp^2\equiv N^IN^J\nabla_I\nabla_JV+\epsilon H^2R_{\rm fs}-\Omega^2,
\qquad R_{\rm fs}=-4/d,
\label{eq:msdef}
\end{equation}
with $N^I$ the unit normal, $\Omega$ the turn rate, and $R_{\rm fs}$ the field-space Ricci scalar. The three terms encode, respectively, Hessian stabilization normal to the path, the correction from field-space curvature, and the loss of decoupling caused by bending. In the $(X,s)$ coordinates one has $G^{XX}=2m^2s^2/d$, so along the valley
\begin{equation}
m_s^2=\frac{4A_pm^2}{d}s^{2-p}+\Order(V_0s^{-q})+\Order(s^{-p-2\nu})+\Order(\epsilon H^2)-\Omega^2 .
\label{eq:ms}
\end{equation}
Since $H^2\to V_0/3$ on the plateau, one obtains
\begin{equation}
\frac{m_s^2}{H^2}=\frac{12A_pm^2}{dV_0}s^{2-p}\big[1+\Order(s^{-q})+\cdots\big].
\label{eq:msoverH}
\end{equation}
Thus, within the branch class specified above, control fails asymptotically for $p>2$, becomes asymptotically strong for $p<2$, and is boundary-sensitive at $p=2$.

The two inequalities have distinct physical meanings. $\Delta>0$ is the plateau condition: the displaced-branch correction must decay faster than the leading saxionic slope. The condition on $p$ is the decoupling condition: the heavy restoring force must not redshift below $H^2$ during the CMB window.

\theorembox{\textbf{Local Control Theorem.} A penumbral valley supports a controlled $q$-universal plateau if and only if $\Delta>0$ and either $p<2$, or $p=2$ with $12A_pm^2/(dV_0)\gg1$ in the inflationary window. Equivalently, penumbral valleys split into three EFT regimes: outside the $q$-universal plateau class ($\Delta\le0$), $q$-plateau without control ($\Delta>0$ and $p>2$), and controlled $q$-plateau.}

The theorem converts valley geometry into an EFT selection rule. A long valley is not enough. The same backreaction that creates the plateau also invalidates it when the quadratic heavy operator softens too quickly: the orthogonal sector then ceases to decouple and the apparent plateau is not a controlled single-clock regime. Existence and control must be imposed together. At the boundary $p=2$, control is supplied by the large prefactor $12A_pm^2/(dV_0)$ across the CMB window. For the $d=1$ benchmark the canonical slope is small, $|U_{,\varphi}|/U=\sqrt{r/8}\simeq1.25\times10^{-2}$, so the model avoids both the late-slope and heavy-decoupling obstructions. Earlier penumbral analyses identify candidate finite-distance windows; the displaced-minimum condition together with the heavy-sector inequality decides whether such a window is already a controlled inflationary EFT.

{\textbf{Flux-compatible Hodge block.}--}
A one-parameter large-complex-structure boundary gives
\begin{equation}
K_{\rm cs}=-\log\!\left(\frac{4\kappa}{3}s^3+\cdots\right),
\qquad 
G_{z\bar z}=\frac{3}{4s^2}+\cdots ,
\label{eq:LCSmetric}
\end{equation}
corresponding to $d=3$ in Eq.~\eqref{eq:metric}; the benchmark below uses the $d=1$ hyperbolic normalization, while all formulae keep $d$ explicit. To keep flux quantization explicit, write the flux bilinear as
\begin{equation}
V_{\rm cs}=\frac{\Lambda_{\rm cs}^4}{2}\rho^T\widehat{\cal Z}_{\rm pen}(s)\rho,
\quad
\rho=(N_0,N_X X)^T,
\quad N_0,N_X\in\mathbb Z,
\label{eq:quantizedrho}
\end{equation}
with
\begin{equation}
\widehat{\cal Z}_{\rm pen}(s)=
\begin{pmatrix}
z_{00}(1-\beta/s)^2 & z_{0X}s^{-3}\\
z_{0X}s^{-3} & z_{XX}s^{-2}
\end{pmatrix}+\cdots .
\label{eq:quantizedHodgeBlock}
\end{equation}
Then
\begin{equation}
V_0=\frac{\Lambda_{\rm cs}^4}{2}z_{00}N_0^2,
\quad
B=\Lambda_{\rm cs}^4z_{0X}N_0N_X,
\quad
A=\frac{\Lambda_{\rm cs}^4}{2}z_{XX}N_X^2,
\label{eq:fluxmap}
\end{equation}
so the block realizes Eq.~\eqref{eq:benchmark} with $q=1$, $p=2$, $\nu=1$, and $\Delta=3>0$. Positivity requires
\begin{equation}
z_{00}z_{XX}s^{-2}(1-\beta/s)^2-z_{0X}^{2}s^{-6}>0 .
\label{eq:quantizedblockpos}
\end{equation}
The benchmark ratios are reproduced by the integer branch choice
\begin{equation}
N_0=2,
\quad
N_X=10,
\quad
z_{0X}=\frac12,
\quad
z_{XX}=1,
\quad
z_{00}=9.80,
\label{eq:integerchoice}
\end{equation}
which gives $B/A=0.2$ and $12A/V_0\simeq30.6$ (hence $12A/(dV_0)\simeq30.6/d$); the observed amplitude fixes only the common scale, $\Lambda_{\rm cs}^4\simeq2\times10^{-12}$. Thus the small CMB-normalized coefficients do not require fractional fluxes: the flux integers fix the ratios, while the compactification prefactor fixes the scale. The nilpotent-orbit derivation, positivity bound, and coefficient map are given in the Supplemental Material\SM.

{\textbf{Analytic Attractor Family.}--}
The Hodge block above motivates the minimal closed-form penumbral realization
\begin{equation}
V_q(a,s)=V_0\Bigl(1-\frac{\beta}{s^q}\Bigr)^2+\frac{A}{s^2}(e-ma)^2+\frac{B}{s^3}(e-ma),
\label{eq:benchmark}
\end{equation}
where $\beta$ sets the leading plateau amplitude, $A$ fixes the heavy restoring force, and $B$ measures the off-diagonal branch displacement. For $q=1$ this is the Hodge block specified by Eqs.~\eqref{eq:quantizedrho}--\eqref{eq:fluxmap}; the parameter $q$ labels the analogous discrete plateau families. For this family,
\begin{align}
X_v(s)&=e-ma_v(s)=-\frac{B}{2A}s^{-1},\\
U_q(s)&=V_0\Bigl(1-\frac{\beta}{s^q}\Bigr)^2-\frac{B^2}{4A}s^{-4}.
\label{eq:exactvalley}
\end{align}
A key technical obstruction in penumbral inflation is that the two-field curved dynamics typically obscure the attractor structure Remarkably, the full curved two-field dynamics reduce exactly to the closed branch equation
\begin{equation}
\partial_{\mathcal N}^2X+\Bigl(3-\epsilon-2\partial_{\mathcal N}\ln s\Bigr)\partial_{\mathcal N}X+\frac{4Am^2}{dH^2}\bigl(X-X_v(s)\bigr)=0,
\label{eq:Xeq}
\end{equation}
where $\mathcal N\equiv\ln a_{\rm FRW}$ denotes e-fold time. This is the \emph{invariant attractor equation}: written directly for the branch variable $X$, it exposes the damped heavy oscillator that locks $X$ to the displaced valley $X_v(s)$ independently of the light-field basis. The attractor is visible inside the full two-field system, not only after asymptotic reduction. The valley is dynamically attractive and adiabatically controlled\SM.

For the cosmic-microwave-background (CMB)-normalized benchmark $(\beta,A,B,m,e)=(1,10^{-10},2\times10^{-11},1,0)$ and $V_0=3.9187\times10^{-11}$, one finds for $d=1$ at $N_*=55$ $(n_s,r)=\bigl(0.96434,1.2493\times10^{-3}\bigr)$, $m_s^2/H^2\simeq30.9$, and $\Omega/H\lesssim1.4\times10^{-7}$, with negligible heavy-turn corrections. The flux-quantized normalization in Eq.~\eqref{eq:fluxmap} reproduces these values with the integer branch data in Eq.~\eqref{eq:integerchoice}; positivity of Eq.~\eqref{eq:quantizedblockpos} is then parametrically safe in the CMB window. The benchmark therefore realizes the boundary-control case $p=2$ with a modest odd-to-even branch hierarchy, while $V_0$ fixes the observed scalar amplitude through the common compactification scale. Varying the reheating history shifts $N_*$ by the usual $\Order(5\text{--}10)$ and moves $(n_s,r)$ along the same narrow corridor; for strict one-parameter LCS normalization $d=3$, $r$ is three times larger and the heavy prefactor is divided by three.

{\textbf{Predictive Stability and Observable Discriminator.}---} Predictive stability requires the control filter to keep higher penumbral corrections from reopening a broad attractor swath. We therefore deform the branch through the complete next penumbral order, including the leading metric correction; the explicit potential, scan ranges, and covariant diagnostics are given in the Supplemental Material\SM. Retaining only models with positive reduced potential, positive kinetic coefficient, $50$--$60$ e-folds, and $m_s^2/H^2>5$ throughout the CMB window, we find for the $d=1$ benchmark at $N_*=55$
\begin{equation}
0.96386\le n_s\le0.96503,
\qquad
1.20\times10^{-3}\le r\le1.29\times10^{-3} .
\label{eq:nlorange}
\end{equation}
The leading running relation is
\begin{equation}
\alpha_s=-\frac{2}{N_*^2}+\Order(N_*^{-3})
=-\frac{q^2}{2d}r+\Order(N_*^{-3}),
\label{eq:alphar}
\end{equation}
so the $d=q=1$ benchmark gives $\alpha_s\simeq-r/2$. This correlated running is the sharpest observable discriminator of the controlled penumbral class, because it ties the tensor amplitude to the same branch data that enforce local control. Oscillatory monodromy remnants of the schematic form $\delta V_{\rm osc}\sim s^{-\gamma}\cos(\omega a+\delta)$ do not spoil the corridor unless they compete already at order $s^{-q}$: along the displaced valley $a_v(s)=a_\infty+\Order(s^{-\nu})$, they only renormalize exponential coefficients at comparable order. Thus control does more than establish existence; it collapses the accepted EFT into a compact bundle in the $(n_s,r)$ and $(\alpha_s,r)$ planes\SM.
\begin{figure}[ht!]
  \includegraphics[width=0.88\columnwidth]{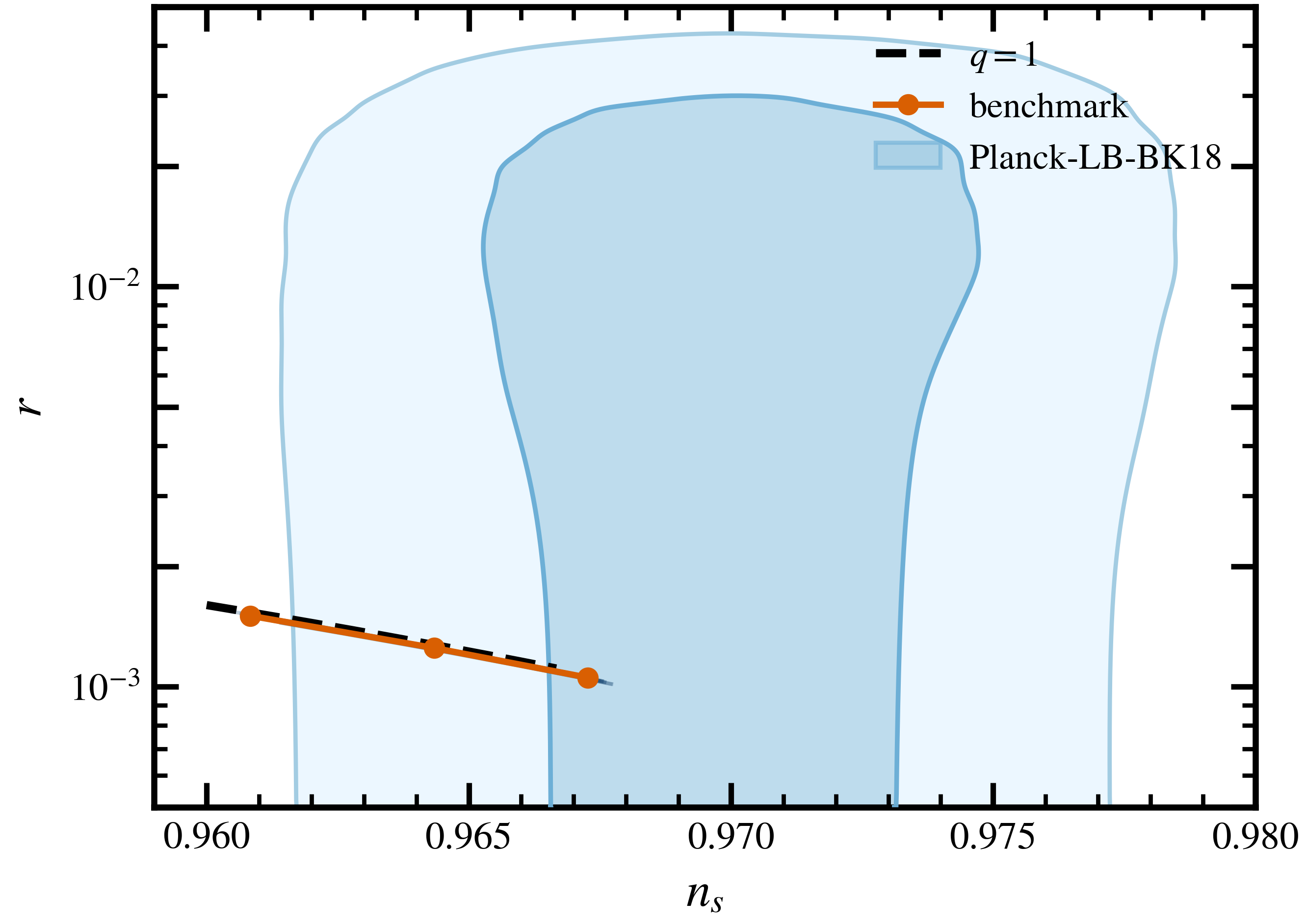}
  \caption{\textbf{Observable corridor.} Controlled next-order deformations remain clustered around the benchmark in the $(n_s,r)$ plane; the same branch data imply the correlated running $\alpha_s\simeq-r/2$ for $d=q=1$.}
  \label{fig:nsr}
\end{figure}

{\textbf{Tameness, Scope, and Remaining Global Input.}---}
The construction lives on a finite-distance penumbral patch where inverse powers of $s$, $\log s$, exponentials of $\log s$, and compact axion phases are tame~\cite{Lanza:2024jxo,Grimm:2021tame,Grimm:2022tameness,Douglas:2024tameness,Lanza:2024ml}. The Hodge-block ansatz shows that the odd branch term is compatible with monodromy, Calabi--Yau power counting at the branch level, and positivity of the retained flux bilinear. Any global completion must still satisfy
\begin{equation}
m_{\rm tower}(\varphi_*)>H_*,
\qquad
\Lambda_{\rm EFT}(\varphi_*)>H_*,
\label{eq:towercontrol}
\end{equation}
together with tadpole cancellation, axio-dilaton stabilization, and K\"ahler stabilization~\cite{Ooguri:2006in,Obied:2018sgi,Garg:2018reu}. Long uplift-like valleys may still fail by steep slope, runaway sectors, or entropy-mode loss~\cite{Lanza:2024jxo,Cicoli:2018tcq,Marchesano:2021gyx}; the theorem isolates the last obstruction locally and decides which valleys merit global embedding. Structurally, hyperbolic geometry supplies the logarithmic field map, while the monodromic branch displacement supplies saxionic rotation, adiabatic control, and the discrete family $\alpha_s\simeq-(q^2/2d)r$~\cite{Carrasco:2015rva,Galante:2014ifa,Kallosh:2015zsa}.

Observationally, the benchmark lies in the concave small-$r$ region favored by Planck plus BICEP/Keck 2018 (BK18) constraints~\cite{Akrami:2018odb,Tristram:2021tvh}, while the predicted tensor scale is targeted by LiteBIRD and future CMB-S4 analyses~\cite{LiteBIRD:2024ghy,Abazajian:2020wmc}. A correlated pattern near $r\sim10^{-3}$ and $\alpha_s\simeq-r/2$ for the $d=q=1$ benchmark would make penumbral inflation sharply testable.

{\textbf{Conclusion.}---}
This work establishes a covariant and falsifiable selection rule for top-down inflationary model building in the penumbra of complex-structure moduli space. Previous penumbral constructions exhibited uplifts and long monodromic valleys but did not produce controlled slow-roll inflation. The essential datum is the branch-displacing odd term: when combined with the entropy-mass criterion, it splits penumbral valleys into outside the $q$-universal class, uncontrolled $q$-plateau, and controlled $q$-plateau. A penumbral construction that realizes this mechanism must exhibit three linked signatures: branch-induced saxionic rotation, the heavy-control inequality, and the associated $(n_s,r,\alpha_s)$ family. The flux-compatible Hodge block above realizes the minimal $(p,\Delta)=(2,3)$ branch data in the branch EFT. While global completion remains the ultimate goal, the computational cost of scanning the flux landscape is prohibitive. Our Local Control Theorem provides a rigorous, necessary EFT filter that eliminates uncontrolled valleys \textit{before} the arduous step of global embedding is attempted, transforming top-down model building from a blind search into a filtered optimization problem. 

{\textbf{Acknowledgement.}---}TL is supported in part by the National Key Research and Development Program of China Grant No. 2020YFC2201504, by the Projects No. 11875062, No. 11947302, No. 12047503, and No. 12275333 supported by the National Natural Science Foundation of China, by the Key Research Program of the Chinese Academy of Sciences, Grant No. XDPB15, by the Scientific Instrument Developing Project of the Chinese Academy of Sciences, Grant No. YJKYYQ20190049, by the International Partnership Program of Chinese Academy of Sciences for Grand Challenges, Grant No. 112311KYSB20210012, and by the Henan Province Outstanding Foreign Scientist Studio Project, No.GZS2025008.

\footnotesize
\bibliography{refs1}

@article{Silverstein:2008sg,
  author = {Silverstein, Eva and Westphal, Alexander},
  journal = {Phys. Rev. D},
  volume = {78},
  pages = {106003},
  year = {2008},
  doi = {10.1103/PhysRevD.78.106003},
  eprint = {0803.3085},
  archivePrefix = {arXiv},
  primaryClass = {hep-th}
}

@article{Ooguri:2006in,
  author = {Ooguri, Hirosi and Vafa, Cumrun},
  journal = {Nucl. Phys. B},
  volume = {766},
  pages = {21--33},
  year = {2007},
  doi = {10.1016/j.nuclphysb.2006.10.033},
  eprint = {hep-th/0605264},
  archivePrefix = {arXiv},
  primaryClass = {hep-th}
}

@misc{Obied:2018sgi,
  author = {Obied, Georges and Ooguri, Hirosi and Spodyneiko, Lev and Vafa, Cumrun},
  year = {2018},
  eprint = {1806.08362},
  archivePrefix = {arXiv},
  primaryClass = {hep-th}
}

@article{Garg:2018reu,
  author = {Garg, Sumit K. and Krishnan, Chethan},
  journal = {JHEP},
  volume = {11},
  pages = {075},
  year = {2019},
  doi = {10.1007/JHEP11(2019)075},
  eprint = {1807.05193},
  archivePrefix = {arXiv},
  primaryClass = {hep-th}
}

@article{McAllister:2008hb,
  author = {McAllister, Liam and Silverstein, Eva and Westphal, Alexander},
  journal = {Phys. Rev. D},
  volume = {82},
  pages = {046003},
  year = {2010},
  doi = {10.1103/PhysRevD.82.046003},
  eprint = {0808.0706},
  archivePrefix = {arXiv},
  primaryClass = {hep-th}
}

@article{Flauger:2009ab,
  author = {Flauger, Raphael and McAllister, Liam and Pajer, Enrico and Westphal, Alexander and Xu, Gang},
  journal = {JCAP},
  volume = {06},
  pages = {009},
  year = {2010},
  doi = {10.1088/1475-7516/2010/06/009},
  eprint = {0907.2916},
  archivePrefix = {arXiv},
  primaryClass = {hep-th}
}

@article{Dong:2010in,
  author = {Dong, Xi and Horn, Bart and Silverstein, Eva and Westphal, Alexander},
  journal = {Phys. Rev. D},
  volume = {84},
  pages = {026011},
  year = {2011},
  doi = {10.1103/PhysRevD.84.026011},
  eprint = {1011.4521},
  archivePrefix = {arXiv},
  primaryClass = {hep-th}
}

@article{Marchesano:2014mla,
  author = {Marchesano, Fernando and Shiu, Gary and Uranga, Angel M.},
  journal = {JHEP},
  volume = {09},
  pages = {184},
  year = {2014},
  doi = {10.1007/JHEP09(2014)184},
  eprint = {1404.3040},
  archivePrefix = {arXiv},
  primaryClass = {hep-th}
}

@article{Blumenhagen:2014nba,
  author = {Blumenhagen, Ralph and Herschmann, Daniela and Plauschinn, Erik},
  journal = {JHEP},
  volume = {01},
  pages = {007},
  year = {2015},
  doi = {10.1007/JHEP01(2015)007},
  eprint = {1409.7075},
  archivePrefix = {arXiv},
  primaryClass = {hep-th}
}

@article{Hebecker:2014eua,
  author = {Hebecker, Arthur and Mangat, Patrick and Rompineve, Fabrizio and Witkowski, Lukas T.},
  journal = {Nucl. Phys. B},
  volume = {894},
  pages = {456--495},
  year = {2015},
  doi = {10.1016/j.nuclphysb.2015.03.015},
  eprint = {1411.2032},
  archivePrefix = {arXiv},
  primaryClass = {hep-th}
}

@article{Lanza:2024jxo,
  author = {Lanza, Stefano and Westphal, Alexander},
  journal = {JHEP},
  volume = {05},
  pages = {071},
  year = {2025},
  doi = {10.1007/JHEP05(2025)071},
  eprint = {2412.12253},
  archivePrefix = {arXiv},
  primaryClass = {hep-th}
}

@article{Achucarro:2012yr,
  author = {Achucarro, Ana and Gong, Jinn-Ouk and Hardeman, Sjoerd and Palma, Gonzalo A. and Patil, Subodh P.},
  journal = {JHEP},
  volume = {05},
  pages = {066},
  year = {2012},
  doi = {10.1007/JHEP05(2012)066},
  eprint = {1201.6342},
  archivePrefix = {arXiv},
  primaryClass = {hep-th}
}

@article{Achucarro:2012sm,
  author = {Achucarro, Ana and Atal, Vicente and Cespedes, Sebastian and Gong, Jinn-Ouk and Palma, Gonzalo A. and Patil, Subodh P.},
  journal = {Phys. Rev. D},
  volume = {86},
  pages = {121301},
  year = {2012},
  doi = {10.1103/PhysRevD.86.121301},
  eprint = {1205.0710},
  archivePrefix = {arXiv},
  primaryClass = {hep-th}
}

@article{Kallosh:2013hoa,
  author = {Kallosh, Renata and Linde, Andrei and Roest, Diederik},
  journal = {JHEP},
  volume = {11},
  pages = {198},
  year = {2013},
  doi = {10.1007/JHEP11(2013)198},
  eprint = {1311.0472},
  archivePrefix = {arXiv},
  primaryClass = {hep-th}
}

@article{Akrami:2018odb,
  author = {Akrami, Yashar and others},
  collaboration = {Planck},
  journal = {Astron. Astrophys.},
  volume = {641},
  pages = {A10},
  year = {2020},
  doi = {10.1051/0004-6361/201833887},
  eprint = {1807.06211},
  archivePrefix = {arXiv},
  primaryClass = {astro-ph.CO}
}

@article{Tristram:2021tvh,
  author = {Tristram, M. and others},
  collaboration = {BICEP/Keck and Planck},
  journal = {Phys. Rev. D},
  volume = {105},
  pages = {083524},
  year = {2022},
  doi = {10.1103/PhysRevD.105.083524},
  eprint = {2112.07961},
  archivePrefix = {arXiv},
  primaryClass = {astro-ph.CO}
}

@article{LiteBIRD:2024ghy,
  author = {Ghigna, T. and others},
  collaboration = {LiteBIRD},
  journal = {Proc. SPIE},
  volume = {13092},
  pages = {1309228},
  year = {2024},
  doi = {10.1117/12.3021377},
  eprint = {2406.02724},
  archivePrefix = {arXiv},
  primaryClass = {astro-ph.IM}
}

@article{Abazajian:2020wmc,
  author = {Abazajian, Kevork N. and others},
  collaboration = {CMB-S4},
  journal = {Astrophys. J.},
  volume = {926},
  pages = {54},
  year = {2022},
  doi = {10.3847/1538-4357/ac1596},
  eprint = {2008.12619},
  archivePrefix = {arXiv},
  primaryClass = {astro-ph.CO}
}

@article{Roest:2015qya,
  author = {Roest, Diederik and Scalisi, Marco},
  journal = {Phys. Rev. D},
  volume = {92},
  pages = {043525},
  year = {2015},
  doi = {10.1103/PhysRevD.92.043525},
  eprint = {1503.07909},
  archivePrefix = {arXiv},
  primaryClass = {hep-th}
}

@article{Senatore:2010wk,
  author = {Senatore, Leonardo and Zaldarriaga, Matias},
  journal = {JHEP},
  volume = {04},
  pages = {024},
  year = {2012},
  doi = {10.1007/JHEP04(2012)024},
  eprint = {1009.2093},
  archivePrefix = {arXiv},
  primaryClass = {hep-th}
}

@article{Landete:2018kft,
  author = {Landete, Aitor and Shiu, Gary},
  journal = {Phys. Rev. D},
  volume = {98},
  pages = {066012},
  year = {2018},
  doi = {10.1103/PhysRevD.98.066012},
  eprint = {1806.01874},
  archivePrefix = {arXiv},
  primaryClass = {hep-th}
}

@article{Grimm:2020cda,
  author = {Grimm, Thomas W. and Li, Chongchuo},
  journal = {JHEP},
  volume = {06},
  pages = {067},
  year = {2021},
  doi = {10.1007/JHEP06(2021)067},
  eprint = {2012.08272},
  archivePrefix = {arXiv},
  primaryClass = {hep-th}
}

@article{Pajer:2024uvs,
  author = {Pajer, Enrico and Wang, Dong-Gang and Zhang, Bowei},
  journal = {SciPost Phys.},
  volume = {20},
  pages = {097},
  year = {2026},
  doi = {10.21468/SciPostPhys.20.4.097},
  eprint = {2412.05762},
  archivePrefix = {arXiv},
  primaryClass = {hep-th}
}

@article{Carrasco:2015rva,
  author = {Carrasco, John Joseph M. and Kallosh, Renata and Linde, Andrei and Roest, Diederik},
  journal = {Phys. Rev. D},
  volume = {92},
  pages = {041301},
  year = {2015},
  doi = {10.1103/PhysRevD.92.041301},
  eprint = {1504.05557},
  archivePrefix = {arXiv},
  primaryClass = {hep-th}
}

@article{Galante:2014ifa,
  author = {Galante, Mario and Kallosh, Renata and Linde, Andrei and Roest, Diederik},
  journal = {Phys. Rev. Lett.},
  volume = {114},
  pages = {141302},
  year = {2015},
  doi = {10.1103/PhysRevLett.114.141302},
  eprint = {1412.3797},
  archivePrefix = {arXiv},
  primaryClass = {hep-th}
}

@article{Baume:2016psm,
  author = {Baume, Florent and Palti, Eran},
  journal = {JHEP},
  volume = {08},
  pages = {043},
  year = {2016},
  doi = {10.1007/JHEP08(2016)043},
  eprint = {1602.06517},
  archivePrefix = {arXiv},
  primaryClass = {hep-th}
}

@article{Valenzuela:2016yny,
  author = {Valenzuela, Irene},
  journal = {JHEP},
  volume = {06},
  pages = {098},
  year = {2017},
  doi = {10.1007/JHEP06(2017)098},
  eprint = {1611.00394},
  archivePrefix = {arXiv},
  primaryClass = {hep-th}
}

@article{Grimm:2021tame,
  author = {Grimm, Thomas W.},
  journal = {JHEP},
  volume = {11},
  pages = {003},
  year = {2022},
  doi = {10.1007/JHEP11(2022)003},
  eprint = {2112.08383},
  archivePrefix = {arXiv},
  primaryClass = {hep-th}
}

@article{Grimm:2022tameness,
  author = {Grimm, Thomas W. and Lanza, Stefano and Li, Chongchuo},
  journal = {JHEP},
  volume = {09},
  pages = {149},
  year = {2022},
  doi = {10.1007/JHEP09(2022)149},
  eprint = {2206.00697},
  archivePrefix = {arXiv},
  primaryClass = {hep-th}
}

@article{Douglas:2024tameness,
  author = {Douglas, Michael R. and Grimm, Thomas W. and Schlechter, Lorenz},
  journal = {Adv. Theor. Math. Phys.},
  volume = {28},
  pages = {2603--2656},
  year = {2024},
  doi = {10.4310/ATMP.241119035402},
  eprint = {2210.10057},
  archivePrefix = {arXiv},
  primaryClass = {hep-th}
}

@article{Lanza:2024ml,
  author = {Lanza, Stefano},
  journal = {Eur. Phys. J. C},
  volume = {84},
  pages = {631},
  year = {2024},
  doi = {10.1140/epjc/s10052-024-12988-z},
  eprint = {2311.03437},
  archivePrefix = {arXiv},
  primaryClass = {hep-th}
}

@article{Kaloper:2008fb,
  author = {Kaloper, Nemanja and Sorbo, Lorenzo},
  journal = {Phys. Rev. Lett.},
  volume = {102},
  pages = {121301},
  year = {2009},
  doi = {10.1103/PhysRevLett.102.121301},
  eprint = {0811.1989},
  archivePrefix = {arXiv},
  primaryClass = {hep-th}
}

@article{Gordon:2000hv,
  author = {Gordon, Christopher and Wands, David and Bassett, Bruce A. and Maartens, Roy},
  journal = {Phys. Rev. D},
  volume = {63},
  pages = {023506},
  year = {2001},
  doi = {10.1103/PhysRevD.63.023506},
  eprint = {astro-ph/0009131},
  archivePrefix = {arXiv},
  primaryClass = {astro-ph}
}

@article{RenauxPetel:2015mga,
  author = {Renaux-Petel, Sebastien and Turzynski, Krzysztof},
  journal = {Phys. Rev. Lett.},
  volume = {117},
  pages = {141301},
  year = {2016},
  doi = {10.1103/PhysRevLett.117.141301},
  eprint = {1510.01281},
  archivePrefix = {arXiv},
  primaryClass = {astro-ph.CO}
}

@article{Hebecker:2017lxm,
  author = {Hebecker, Arthur and Henkenjohann, Philipp and Witkowski, Lukas T.},
  journal = {JHEP},
  volume = {12},
  pages = {033},
  year = {2017},
  doi = {10.1007/JHEP12(2017)033},
  eprint = {1708.06761},
  archivePrefix = {arXiv},
  primaryClass = {hep-th}
}

@article{Grimm:2018ohb,
  author = {Grimm, Thomas W. and Palti, Eran and Valenzuela, Irene},
  journal = {JHEP},
  volume = {08},
  pages = {143},
  year = {2018},
  doi = {10.1007/JHEP08(2018)143},
  eprint = {1802.08264},
  archivePrefix = {arXiv},
  primaryClass = {hep-th}
}

@article{Lanza:2021qsu,
  author = {Lanza, Stefano and Marchesano, Fernando and Martucci, Luca and Valenzuela, Irene},
  journal = {JHEP},
  volume = {09},
  pages = {197},
  year = {2021},
  doi = {10.1007/JHEP09(2021)197},
  eprint = {2104.05726},
  archivePrefix = {arXiv},
  primaryClass = {hep-th}
}

@article{Marchesano:2021gyx,
  author = {Marchesano, Fernando and Prieto, David and Wiesner, Max},
  journal = {JHEP},
  volume = {08},
  pages = {077},
  year = {2021},
  doi = {10.1007/JHEP08(2021)077},
  eprint = {2105.09326},
  archivePrefix = {arXiv},
  primaryClass = {hep-th}
}

@article{Cicoli:2018tcq,
  author = {Cicoli, Michele and Ciupke, David and Mayrhofer, Christoph and Shukla, Pramod},
  journal = {JHEP},
  volume = {05},
  pages = {001},
  year = {2018},
  doi = {10.1007/JHEP05(2018)001},
  eprint = {1801.05434},
  archivePrefix = {arXiv},
  primaryClass = {hep-th}
}

@article{Kallosh:2015zsa,
  author = {Kallosh, Renata and Linde, Andrei},
  journal = {Comptes Rendus Physique},
  volume = {16},
  pages = {914--927},
  year = {2015},
  doi = {10.1016/j.crhy.2015.07.004},
  eprint = {1503.06785},
  archivePrefix = {arXiv},
  primaryClass = {hep-th}
}

\clearpage
\renewcommand{\theequation}{A\arabic{equation}}
\setcounter{equation}{0}
\onecolumngrid
\appendix
\begin{center}
  \textbf{\large Supplementary Material}\\[.2cm]
\end{center}

This Supplemental Material gives the derivations and diagnostics underlying the Letter. Section~I derives the local universality class and heavy-control filter. Section~II gives the covariant entropy-mass derivation. Section~III develops the closed-form benchmark family and invariant attractor equation. Section~IV treats the complete next penumbral order and the documented deformation scan. Section~V derives the concrete one-parameter Calabi--Yau Hodge block used in the main text. Section~VI fixes the broader Type-IIB origin, scope of the theorem, and remaining diagnostics.

\section{Local reduction and the universal plateau}

The local monodromy-preserving penumbral EFT used in the Letter is
\begin{equation}
V(X,s)=V_0-c_qV_0s^{-q}+\sum_{n>q}\frac{v_n}{s^n}
+\sum_{n\ge p}\frac{A_n}{s^n}X^2
+\sum_{n\ge p+\nu}\frac{B_n}{s^n}X+\Order(X^3),
\end{equation}
with hyperbolic metric $G_{IJ}=\delta_{IJ}/(2s^2)$ and $X\equiv e-ma$. Minimization with respect to $X$ gives
\begin{equation}
\partial_XV=2A_pXs^{-p}+B_{p+\nu}s^{-(p+\nu)}+\cdots=0,
\end{equation}
and therefore
\begin{equation}
X_v(s)=-\frac{B_{p+\nu}}{2A_p}s^{-\nu}+\Order(s^{-\nu-1}).
\end{equation}
Since $a=(e-X)/m$, the valley slope is
\begin{equation}
a_v'(s)=-\frac{X_v'(s)}{m}=\Order(s^{-\nu-1}),
\end{equation}
so the induced line element along the valley is
\begin{equation}
d\sigma^2=\frac{1+a_v'(s)^2}{2s^2}ds^2
=\frac{1}{2s^2}\left[1+\Order(s^{-2\nu-2})\right]ds^2,
\end{equation}
which integrates to
\begin{equation}
\varphi=\frac{\log s}{\sqrt2}+\Order(s^{-2\nu-2}).
\end{equation}
Substituting the valley back into the local potential yields
\begin{equation}
U(s)=V_0-c_qV_0s^{-q}-\frac{B_{p+\nu}^2}{4A_p}s^{-(p+2\nu)}+\cdots,
\end{equation}
so the canonical potential becomes
\begin{equation}
U(\varphi)=V_0-c_qV_0\ee^{-q\sqrt2\varphi}
-\frac{B_{p+\nu}^2}{4A_p}\ee^{-(p+2\nu)\sqrt2\varphi}+\cdots.
\end{equation}
Whenever $\Delta\equiv p+2\nu-q>0$, the heavy-branch correction is asymptotically subleading and the leading potential is a plateau determined entirely by $q$. The e-fold integral then gives, with $N_*$ the number of e-folds before the end of inflation,
\begin{equation}
N_*\simeq\int\frac{U}{U_{,\varphi}}\,d\varphi
\simeq\frac{\ee^{q\sqrt2\varphi_*}}{2q^2c_q},
\end{equation}
so that
\begin{equation}
\epsilon_*\simeq\frac{1}{4q^2N_*^2},
\qquad
\eta_*\simeq-\frac{1}{N_*},
\end{equation}
and therefore
\begin{equation}
n_s=1-\frac{2}{N_*}+\Order(N_*^{-2}),
\qquad
r=\frac{4}{q^2N_*^2}+\Order(N_*^{-3}).
\end{equation}

\section{Covariant entropy mass and the heavy-control filter}

In the $(X,s)$ variables the field-space metric is
\begin{equation}
ds_{\rm field}^2=\frac{1}{2s^2}\left(\frac{dX^2}{m^2}+ds^2\right).
\end{equation}
Let $y(s)\equiv X_v'(s)/m$. The orthonormal tangent and normal vectors are
\begin{equation}
T^I=\frac{\sqrt2s}{\sqrt{1+y^2}}(my,1),
\qquad
N^I=\frac{\sqrt2s}{\sqrt{1+y^2}}(m,-y),
\end{equation}
with $G_{IJ}T^IT^J=G_{IJ}N^IN^J=1$ and $G_{IJ}T^IN^J=0$. The covariant entropy mass is
\begin{equation}
m_s^2=N^IN^J\nabla_I\nabla_JV+\epsilon H^2R_{\rm fs}-\Omega^2,
\end{equation}
with $R_{\rm fs}=-4$. Because $X_v'(s)=\Order(s^{-\nu-1})$, the normal vector aligns asymptotically with the $X$ direction. Using $\nabla_X\nabla_XV=2A_ps^{-p}+\cdots$ and $G^{XX}=2m^2s^2$, one obtains
\begin{equation}
m_s^2=4A_pm^2s^{2-p}\left[1+\Order(s^{-2\nu-2})\right]
+\Order(s^{-p-2\nu})+\Order(\epsilon H^2).
\end{equation}
Since $H^2\to V_0/3$ on the plateau, one has
\begin{equation}
\frac{m_s^2}{H^2}
=
\frac{12A_pm^2}{V_0}s^{2-p}\left[1+\Order(s^{-2\nu-2})\right].
\end{equation}
Thus asymptotic control is automatic for $p<2$, fails for $p>2$, and is boundary-sensitive at $p=2$, where one requires the prefactor $12A_pm^2/V_0$ to be large in the inflationary window. The turn rate also decays. For small valley slope, the geodesic curvature obeys $k_g\sim s^{-\nu-1}$, while on a $q$-plateau one has $\epsilon\sim s^{-2q}$. Therefore
\begin{equation}
\frac{\Omega}{H}=k_g\sqrt{2\epsilon}
=\Order\!\left(s^{-q-\nu-1}\right),
\end{equation}
so heavy-turn corrections are strongly suppressed in the controlled region.

\section{Closed-form benchmark family and invariant attractor equation}

The minimal closed-form family is
\begin{equation}
V_q(a,s)=V_0\Bigl(1-\frac{\beta}{s^q}\Bigr)^2
+\frac{A}{s^2}(e-ma)^2+\frac{B}{s^3}(e-ma).
\end{equation}
Its heavy valley is analytic,
\begin{equation}
a_v(s)=\frac{e}{m}+\frac{B}{2Am}\frac{1}{s},
\qquad
X_v(s)=-\frac{B}{2A}\frac{1}{s},
\end{equation}
and the reduced potential is
\begin{equation}
U_q(s)=V_0\Bigl(1-\frac{\beta}{s^q}\Bigr)^2-\frac{B^2}{4A}\frac{1}{s^4}.
\end{equation}
For $q=1$ this becomes
\begin{equation}
U(\varphi)=V_0\Bigl(1-\beta\ee^{-\sqrt2\varphi}\Bigr)^2
-\frac{B^2}{4A}\ee^{-4\sqrt2\varphi}.
\end{equation}
The full background equations admit a closed-form rewriting in the invariant coordinate,
\begin{equation}
\partial_{\mathcal N}^2X
+\Bigl(3-\epsilon-2\partial_{\mathcal N}\ln s\Bigr)\partial_{\mathcal N}X
+\frac{4Am^2}{H^2}\bigl(X-X_v(s)\bigr)=0,
\end{equation}
where $\mathcal N\equiv\ln a_{\rm FRW}$ denotes e-fold time and
\begin{equation}
\epsilon=\frac{(\partial_{\mathcal N}a)^2+(\partial_{\mathcal N}s)^2}{4s^2}.
\end{equation}
In terms of the canonical heavy fluctuation $X=X_v+\sqrt2ms\chi$, one finds
\begin{equation}
m_s^2=4Am^2+\Order(s^{-4}),
\end{equation}
which realizes the saturation case $p=2$.

For the benchmark values
\begin{equation}
(\beta,A,B,m,e)=\left(1,10^{-10},2\times10^{-11},1,0\right),
\qquad
V_0=3.9187164325\times10^{-11},
\end{equation}
the reduced single-field valley EFT gives
\begin{align}
N_*=50:&\quad (n_s,r)=(0.96083,1.5046\times10^{-3}),\\
N_*=55:&\quad (n_s,r)=(0.96434,1.2493\times10^{-3}),\\
N_*=60:&\quad (n_s,r)=(0.96727,1.0540\times10^{-3})
\end{align}
with $m_s^2/H^2\simeq 30.9$ across the CMB window and $\Omega/H\lesssim 1.4\times10^{-7}$.
The benchmark requires only a modest hierarchy within the local flux expansion. The ratio $B/A=0.2$ fixes the odd-to-even branch weight, $V_0$ fixes the observed scalar amplitude, and in a compact embedding the allowed coefficient ratios are discretized by flux quantization rather than scanned as arbitrary continuous numbers.
Full two-field integrations rapidly approach the analytic valley and produce only small numerical shifts, consistent with adiabatic decoupling and the negligible turn rate.

\section{Complete next penumbral order and documented scan}

The complete local monodromy-preserving EFT through the next penumbral order, still truncated at quadratic order in $X$, is
\begin{equation}
V_{\rm NLO}=V_0\Bigl(1-\frac{\beta}{s}+\frac{\lambda}{s^2}\Bigr)^2
+
\Bigl(\frac{A}{s^2}+\frac{D}{s^3}\Bigr)X^2
+
\Bigl(\frac{B}{s^3}+\frac{C}{s^4}\Bigr)X,
\end{equation}
with the leading saxionic metric deformation
\begin{equation}
ds_{\rm field}^2=\frac{1}{2s^2}\Bigl[(1+\gamma_a/s)\,da^2+(1+\gamma_s/s)\,ds^2\Bigr].
\end{equation}
The heavy branch remains analytic,
\begin{equation}
X_v(s)=-\frac{Bs+C}{2s(As+D)}
=-\frac{B}{2A}s^{-1}
+\Bigl(-\frac{C}{2A}+\frac{BD}{2A^2}\Bigr)s^{-2}
+\Order(s^{-3}),
\end{equation}
and the reduced potential is
\begin{equation}
U_{\rm NLO}(s)=V_0\Bigl(1-\frac{\beta}{s}+\frac{\lambda}{s^2}\Bigr)^2
-\frac{(Bs+C)^2}{4s^5(As+D)}.
\end{equation}
Thus the leading plateau coefficient is unchanged. The effective kinetic map also stays logarithmic at leading order,
\begin{equation}
\frac{d\varphi}{ds}=\frac{1}{\sqrt2s}\left(1+\frac{\gamma_s}{2s}+\Order(s^{-2})\right),
\end{equation}
so metric deformations move only subleading coefficients in the exponentiated potential.

We scan the dimensionless deformations
\begin{equation}
\hat\lambda\in[-0.35,0.35],\qquad
\hat C\in[-1,1],\qquad
\hat D\in[-0.55,0.55],\qquad
\hat\gamma_s\in[-0.8,0.8],
\end{equation}
where
\begin{equation}
\hat\lambda\equiv\frac{\lambda}{\beta^2},\qquad
\hat C\equiv\frac{C}{Bs_p},\qquad
\hat D\equiv\frac{D}{As_p},\qquad
s_p=225.
\end{equation}
In the scan shown here we set $\gamma_a=0$, since it only renormalizes the heavy-sector kinetic normalization and does not modify the leading plateau data. Models are retained only if the reduced potential is positive, the effective kinetic coefficient is positive, $50$--$60$ e-fold inflation exists, and $m_s^2/H^2>5$ throughout the CMB window. Among the accepted models we obtain at $N_*=55$
\begin{equation}
0.96386\le n_s\le0.96503,\qquad
1.20\times10^{-3}\le r\le1.29\times10^{-3},
\end{equation}
with representative control values $m_s^2/H^2\gtrsim 34$ and $\Omega/H\lesssim 10^{-11}$. The same accepted set gives a narrow running corridor centered on the leading $q=1$ relation $\alpha_s\simeq-r/2$, which is why the main-text observational bundle remains compact once the local EFT constraints are imposed.

\begin{figure}[t]
  \includegraphics[width=0.96\textwidth]{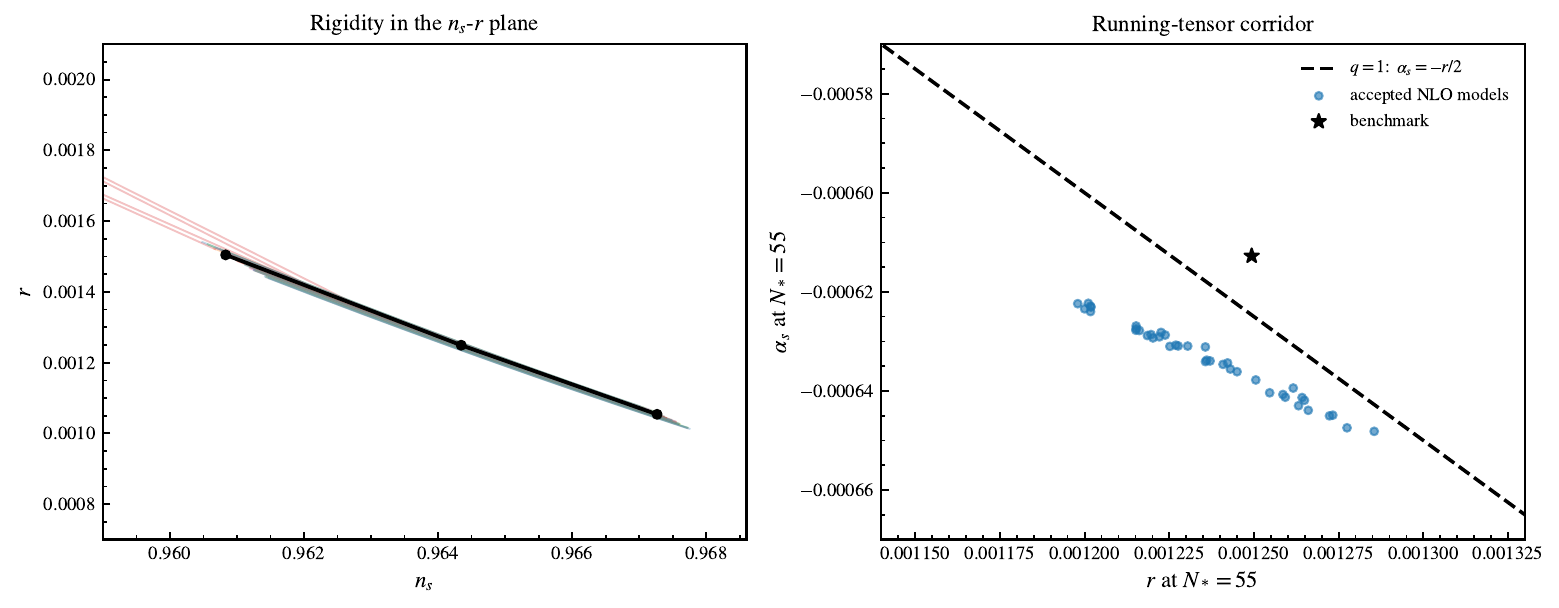}
  \caption{\textbf{Predictive stability and observational corridor.} (a) Slow-roll tracks for the complete next-order monodromy-preserving EFT and mild metric deformations; the benchmark is the thick black curve. (b) The same models in the $(r,\alpha_s)$ plane compared with the leading $q=1$ relation $\alpha_s=-r/2$. After control is imposed, the local EFT predicts a compact observational corridor rather than a broad attractor swath.}
  \label{fig:s2}
\end{figure}

\begin{figure}[t]
  \includegraphics[width=0.96\textwidth]{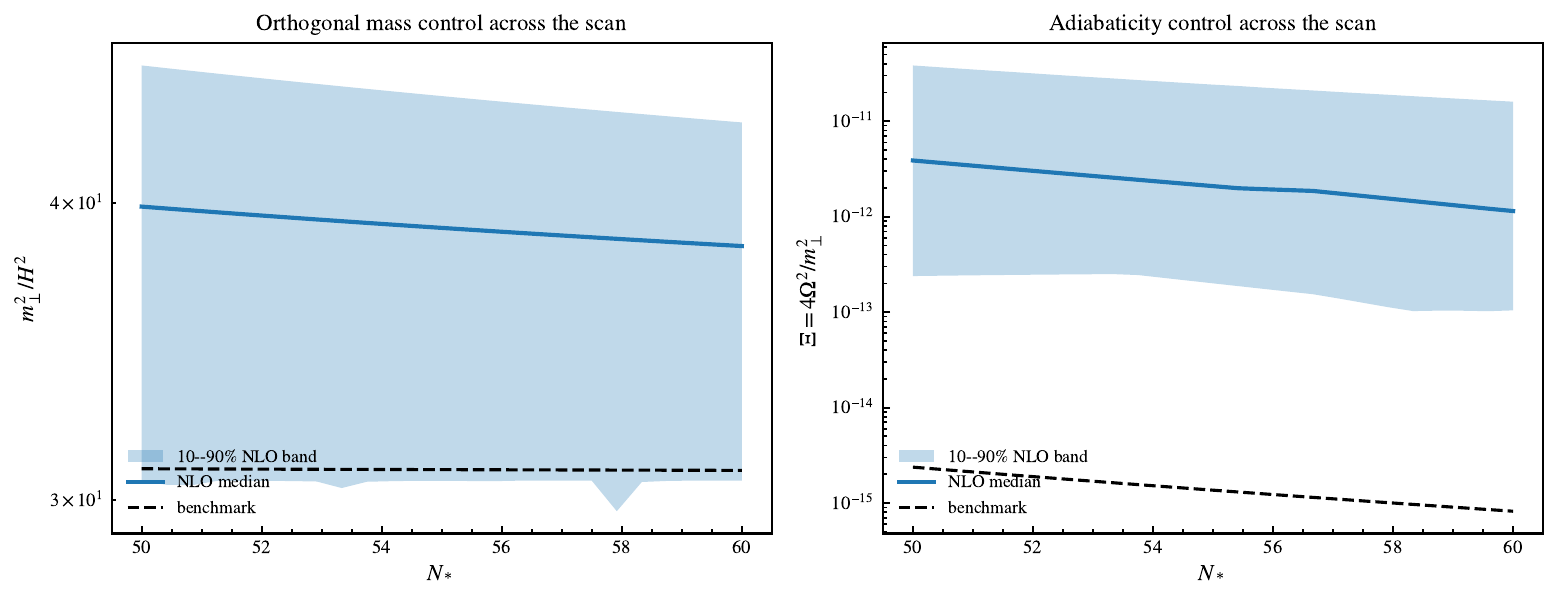}
  \caption{\textbf{Control envelopes for the accepted scan.} The upper envelopes of $\Omega/H$ remain tiny throughout the CMB window.}
  \label{fig:s3}
\end{figure}

\section{Concrete one-parameter Calabi--Yau Hodge block}

This section gives the derivation of the local Calabi--Yau block used in the main text. Near a one-parameter large-complex-structure boundary, the nilpotent orbit gives a period vector of the schematic form
\begin{equation}
\Pi_{\rm nil}(z)=
\left(
1,\,
z,\,
\frac{\kappa}{2}z^2+\cdots,\,
-\frac{\kappa}{6}z^3+\cdots
\right)^T ,
\qquad z=a+is ,
\end{equation}
where $\kappa>0$ is the classical triple-intersection number and the omitted terms are lower-degree polynomial terms and exponentially small instanton corrections. The complex-structure K\"ahler potential is
\begin{equation}
K_{\rm cs}=-\log\!\left(\frac{4\kappa}{3}s^3+\cdots\right),
\end{equation}
so that
\begin{equation}
G_{z\bar z}=\partial_z\partial_{\bar z}K_{\rm cs}
=
\frac{3}{4s^2}+\Order(s^{-5}) .
\end{equation}
After a constant local normalization of the saxion, this is the hyperbolic metric used in the main text.

The flux-induced complex-structure potential can be written locally as a positive Hodge bilinear
\begin{equation}
V_{\rm cs}=\frac12\,\rho^T{\cal Z}(s)\rho ,
\end{equation}
where the components of $\rho$ are monodromy-invariant axion polynomials. On a chosen branch take the two-component subspace
\begin{equation}
\rho=(\rho_0,\rho_X)^T=(1,X)^T,
\qquad X=e-ma .
\end{equation}
The most economical penumbral block that contains an uplift, a quadratic branch-restoring term, and an odd branch displacement is
\begin{equation}
{\cal Z}_{\rm pen}(s)=
\begin{pmatrix}
2V_0(1-\beta/s)^2 & B s^{-3}\\
B s^{-3} & 2A s^{-2}
\end{pmatrix}
+\Order(s^{-3})_{00}
+\Order(s^{-4})_{0X}
+\Order(s^{-3})_{XX}.
\end{equation}
Substituting this block into the bilinear gives
\begin{equation}
V_{\rm cs}
=
V_0\left(1-\frac{\beta}{s}\right)^2
+\frac{A}{s^2}X^2
+\frac{B}{s^3}X+\cdots .
\end{equation}
Thus the block realizes
\begin{equation}
q=1,\qquad p=2,\qquad \nu=1,\qquad \Delta=p+2\nu-q=3>0 .
\end{equation}
The heavy branch is
\begin{equation}
X_v(s)=-\frac{B}{2A}\frac1s ,
\end{equation}
and the reduced potential is
\begin{equation}
U(s)=V_0\left(1-\frac{\beta}{s}\right)^2-\frac{B^2}{4A}s^{-4}+\cdots .
\end{equation}
The retained Hodge block is positive whenever
\begin{equation}
2V_0(1-\beta/s)^2>0,
\qquad
2As^{-2}>0,
\end{equation}
and
\begin{equation}
\det{\cal Z}_{\rm pen}
=
4AV_0s^{-2}(1-\beta/s)^2-B^2s^{-6}>0 .
\end{equation}
Equivalently,
\begin{equation}
\frac{B^2}{4AV_0s^4(1-\beta/s)^2}<1 .
\end{equation}
For the benchmark
\begin{equation}
(\beta,A,B,m,e)=\left(1,10^{-10},2\times10^{-11},1,0\right),
\qquad
V_0=3.9187164325\times10^{-11},
\end{equation}
and $s=\Order(10^2)$ in the CMB window, this ratio is many orders of magnitude below unity. The odd branch term is therefore compatible with positivity of the retained Hodge bilinear.

The entropy-mass condition gives
\begin{equation}
\frac{m_s^2}{H^2}=\frac{12Am^2}{V_0}+\Order(s^{-2})
\simeq 30.6 ,
\end{equation}
so the same block realizes the boundary-control case $p=2$ with a large heavy prefactor. The example is local in the precise sense used in the Letter: it specifies a Calabi--Yau large-complex-structure Hodge block and the monodromy-invariant branch data. A full compact embedding must still realize the block with integral fluxes and satisfy tadpole cancellation, axio-dilaton stabilization, K\"ahler stabilization, and tower control.

\section{String-theoretic origin, scope, and additional diagnostics}

\noindent\textbf{Type-IIB origin of the local ansatz.}
Near a one-parameter Calabi--Yau complex-structure boundary the period vector is controlled by a nilpotent-orbit expansion, so the prepotential and periods are polynomial in the local modulus $z=a+is$ up to exponentially small corrections. After fixing the axio-dilaton and using the no-scale complex-structure potential, one obtains a scalar potential built from inverse powers of $s$ multiplying monodromy-invariant axion polynomials~\cite{Grimm:2020cda,Grimm:2021tame,Grimm:2022tameness,Lanza:2024ml}. A convenient way to state the general local structure is
\begin{equation}
V_{\rm flux}(a,s)=\mathcal G_{IJ}(s)\,\rho^I(a)\rho^J(a),
\qquad
\mathcal G_{IJ}(s)\sim s^{-n_{IJ}},
\end{equation}
where the $\rho^I$ are invariant under the axionic monodromy action on a chosen branch. Expanding those polynomials around a local branch coordinate $X\equiv e-ma$ gives
\begin{equation}
\rho^I(X)=\rho_0^I+\rho_1^I X+\rho_2^I X^2+\cdots,
\end{equation}
and therefore
\begin{equation}
V_{\rm flux}(X,s)=V_{\rm const}(s)+B(s)X+A(s)X^2+\Order(X^3),
\end{equation}
with
\begin{equation}
V_{\rm const}(s)=\mathcal G_{IJ}\rho_0^I\rho_0^J,\qquad
B(s)=2\mathcal G_{IJ}\rho_0^I\rho_1^J,\qquad
A(s)=\mathcal G_{IJ}\rho_1^I\rho_1^J+\cdots.
\end{equation}
Because the matrix $\mathcal G_{IJ}(s)$ carries the asymptotic saxionic weights, this local branch expansion generically yields an $X$-independent sector together with linear and quadratic heavy terms whose coefficients scale as inverse powers of $s$. Equation~(A1) of the Letter is therefore the natural local branch expansion of a Type-IIB flux potential in the penumbral regime. The odd term appears whenever the chosen branch has a nonzero invariant offset $\rho_0^I$ and a nonzero linear monodromy component $\rho_1^I$; monodromy permits it, while symmetry or flux tuning may remove it.

\noindent\textbf{What remains for an explicit compactification.}
The derivation above establishes the \emph{structure} of the local EFT. A full top-down embedding still requires a concrete Calabi--Yau orientifold, a flux choice, stabilization of the axio-dilaton and K\"ahler sector, and an explicit extraction of the coefficients $(V_0,c_q,A_p,B_{p+\nu})$. The local theorem controls adiabatic decoupling of the orthogonal mode inside the complex-structure EFT; the full Distance-Conjecture tower remains an additional compactification-level constraint. Any explicit embedding must therefore satisfy
\begin{equation}
m_{\rm tower}(\varphi)\gg H(\varphi),
\qquad
\Lambda_{\rm EFT}(\varphi)\gg H(\varphi),
\end{equation}
throughout the CMB window. The scope of the Letter is deliberately local: before global completion is attempted, the penumbral branch data already decide whether a given valley is a controlled inflationary EFT worth pursuing.

\noindent\textbf{Tameness,  and the scope of the inflationary claim.}
The Letter isolates a finite-distance inflationary window inside the penumbra, not an inflationary solution in the strict asymptotic regime. This finite-distance scope is compatible with the tame-geometry logic emphasized in Refs.~\cite{Grimm:2021tame,Grimm:2022tameness,Douglas:2024tameness}. On a penumbral patch
\begin{equation}
U=\{|a|<1,\ s>s_0\},
\end{equation}
the local ingredients used in the Letter are definable in the same o-minimal structures relevant for complex-structure sectors: inverse powers of $s$, logarithms, exponentials of $\log s$, restricted analytic functions, and periodic functions of the compact axion. In particular,
\begin{equation}
s^{-n}=\exp[-n\log s]
\end{equation}
is tame on $s>s_0>1$, while $\cos(\omega a+\delta)$ is tame because $a$ remains compact. Ref.~\cite{Lanza:2024jxo} identified long uplift-like penumbral valleys and clarified why such valleys need not yet yield full slow-roll inflationary solutions. The present Letter sharpens that analysis by identifying the displaced-minimum criterion and heavy-sector inequality that decide when a finite-distance penumbral window is already a \emph{controlled} local inflationary EFT.

\begin{figure}[t]
  \includegraphics[width=0.96\textwidth]{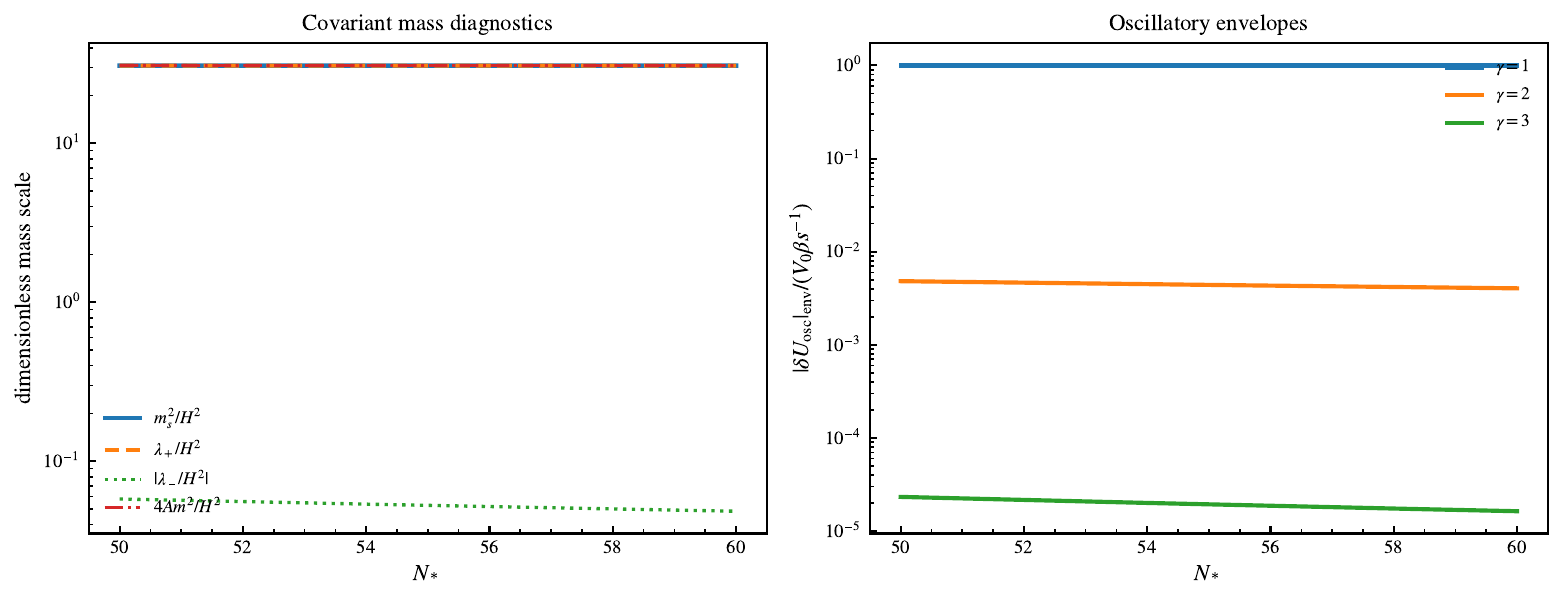}
  \caption{\textbf{Covariant entropy-mass diagnostics and oscillatory envelopes.} (a) $m_s^2/H^2$, the heavy eigenvalue $\lambda_+/H^2$, the light eigenvalue, and the shortcut approximation $4Am^2/H^2$. (b) Representative oscillatory envelopes normalized to the leading plateau term.}
  \label{fig:s4}
\end{figure}

The Letter uses the covariant entropy mass rather than a shortcut mass because the field space is curved. Figure~\ref{fig:s4} shows that the projected entropy mass tracks the heavy eigenvalue of the full covariant mass matrix. The same figure also shows representative oscillatory envelopes along the valley. For corrections of the form $\delta U_{\rm osc}\propto s^{-\gamma}\cos(\omega a+\delta)$, the cases $\gamma>1$ are parametrically subleading to the leading $q=1$ plateau term, while $\gamma=1$ only renormalizes the coefficient of the leading exponential.

\end{document}